\documentstyle[aps,prb,multicol,epsf,graphicx]{revtex}

\newcommand \bea {\begin{eqnarray}}
\newcommand \eea {\end{eqnarray}}
\newcommand \be {\begin{equation}}
\newcommand \ee {\end{equation}}

\newcommand \bi {\bibitem}
\newcommand \s {\sigma}

\begin{document}

\title{Statistics of lowest excitations in two dimensional Gaussian
spin glasses} 

\author{M. Picco$^{(1)}$, F. Ritort$^{(1,2)}$ and M. Sales$^{(2)}$} 

\address{(1) LPTHE, Universit\'e Pierre et Marie Curie, Paris VI
        et Universit\'e Denis Diderot, Paris VII\\
        Boite 126, Tour 16, 1$^{\it er}$ \'etage, 4 place Jussieu,
        F-75252 Paris Cedex 05, France\\
(2) Departament de F\'{\i}sica Fonamental,
 Facultat de F\'{\i}sica, Universitat de Barcelona\\ Diagonal 647,
 08028 Barcelona,Spain
}


\maketitle

\begin{abstract}
A detailed investigation of lowest excitations in two-dimensional
Gaussian spin glasses is presented. We show the existence of a new
zero-temperature exponent $\lambda$ describing the relative number of
finite-volume excitations with respect to large-scale ones. This exponent
yields the standard thermal exponent of droplet theory $\theta$
through the relation, $\theta=d(\lambda-1)$. Our work provides a new
way to measure the thermal exponent $\theta$ without any assumption
about the procedure to generate typical low-lying excitations. We find
clear evidence that $\theta<\theta_{DW}$ where $\theta_{DW}$ is the
thermal exponent obtained in domain-wall theory showing that MacMillan
excitations are not typical.
\end{abstract}

\begin{multicols}{2}
\narrowtext Despite three decades of work in the field of spin glasses
major issues such as their low-temperature behavior still remain
unresolved \cite{REVIEW}. One of the main achievements has been a good
understanding of mean-field theory \cite{MFT} which nevertheless does
not include spatial effects which manifest as droplet
excitations. Leaving aside the controversy whether replica symmetry
breaking is or not a good description of the spin-glass phase
\cite{RSB}, it is reasonable to expect that a phenomenological
approach to the spatial structure of lowest excitations should
satisfactorily account for their low-temperature properties. McMillan
proposed \cite{MILLAN1} that thermal properties in spin glasses are
determined by the scaling behavior of the typical largest excitations
present in the system. Therefore, he assumed that the energy cost of
large scale excitations of length $L$ scales like $L^{\theta}$,
$\theta$ being the thermal exponent. Using domain-wall renormalization
group ideas he also introduced a practical way to determine the
leading energy cost of low-lying large-scale excitations
\cite{MILLAN2}. It consisted in measuring the energy defect of a
domain-wall spanning the whole system obtained by computing the change
of the ground state energy when switching from periodic to anti
periodic boundary conditions in one direction.  This idea has been
further elaborated and extended to deal with equilibrium and dynamical
properties of spin glasses in a scenario nowadays referred as droplet
model \cite{DROPLET}. Several works have used McMillan's method to
determine the value of $\theta$ in two and three dimensions
\cite{VARIOS}.

The purpose of this work is to show an alternative approach to
determine the low $T$ behavior of spin glasses by studying the size
and energy spectrum of the lowest excitations by introducing two
exponents ($\lambda$ and $\theta'$) needed to fully characterize the
zero-temperature fixed point. The {\em entropic} exponent $\lambda$
describes the probability to find a large-scale lowest excitation
spanning the whole system, while the exponent $\theta'$ describes the
system-size dependence of the energy cost of this excitation.

The underlying theoretical background of our approach is the
following. To investigate the leading low-temperature behavior in spin
glasses let us consider expectation values for moments of the order
parameter by keeping only the ground state and the first
excitation. This approach was introduced in \cite{RS} and can be
extended to higher-order excitations to construct a low-temperature
expansion for spin glasses. This is described in detail in a separate
publication \cite{NEXT}. For simplicity, here we restrict the analysis
to the first linear expansion in $T$ by keeping only the first
excitation. If $q=\lbrace\sigma,\tau\rbrace$ denotes the overlap
between two replicas (i.e. configurations of different systems with
the same realization of quenched disorder), then the expectation value
$\overline{\langle q^2\rangle}$ can be written as follows \cite{RS},

\be
\overline{<q^2>}=1-\frac{2}{V^2}\sum_v\int_0^{\infty}dEP(v,E)v(V-v){\rm sech}^2\bigl(\frac{E}{2T}\bigr),
\label{eq1}
\ee

\noindent
where $P(v,E)$ is the joint probability distribution to find a
sample (among an ensemble of ${\cal N}_s$ samples) where the lowest
excitation has $v$ spins overturned respect to the ground state (so
the overlap between the ground and that excited state is $q=1-2v/V$,
$V$ being the total volume of the system) and with energy cost or gap
$E$. If $v_s$ and $E(s)$ denote the volume and
excitation energy of the lowest excitation for sample $s$ then,

\be
P(v,E)=\frac{1}{{\cal N}_s}\sum_{s=1}^{{\cal
N}_s}\delta(v-v_s)\delta(E-E(s))~~.
\label{eq2}
\ee

\noindent
Using the Bayes theorem this joint probability distribution can be
written as $P(v,E)=g_v\hat{P}_v(E)$ where
$\sum_{v=1}^{\frac{V}{2}}g_v=\int_0^{\infty}dE\hat{P}_v(E)=1$, the last equality being valid for any
$v$. $g_v$ is the probability to find a sample such that its lowest
excitation has volume $v$ and $\hat{P}_v(E)$ is the conditioned
probability for this excitation to have a gap equal to $E$.  A
simple low-temperature expansion of (\ref{eq1}) \cite{RS,NEXT} up to
linear order in $T$ yields,

\be
\overline{<q^2>}=1-\frac{4T}{V^2}\sum_{v=1}^V\,g_v\hat{P}_v(0)v(V-v)
\label{eq3}
\ee

\noindent
showing that the leading behavior is determined by both the $g_v$ and
the density of states at zero gap $\hat{P}_v(0)$. In the droplet model
it is generally assumed that typical lowest excitations have average
volume $\overline v=\sum_v v g_v$ diverging with $V=L^D$ and density
weight at zero gap $\hat{P}_V(0)\sim 1/L^{\theta}$ where $\theta$ is
the thermal exponent. In principle, a single exponent $\theta$
describes the scaling behavior of large-scale excitations with volume
$v\propto V$ which are supposed to determine the zero-temperature
critical behavior. Actually, assuming that $g_1/g_V\sim O(1)$,
eq.(\ref{eq3}) suggests that the contribution of small-scale
excitations to the sum in the r.h.s would be negligible in the large
$V$ limit.  Under these reasonable assumptions one obtains
$\overline{<q^2>}=1-cT/L^{\theta}$, where $c$ is a non-universal
stiffness constant related to the particular model. Our purpose here
is to show that $g_1/g_V$ diverges in the large $V$ limit in such a way
that the contribution of small-scale excitations to the r.h.s of
(\ref{eq3}) is as important as the contribution of the large-scale
ones, thus leading to a new physical meaning of the thermal exponent
$\theta$.
\begin{figure}[tbp]
\begin{center}
\rotatebox{270}{
\includegraphics*[width=5cm,height=8.5cm]{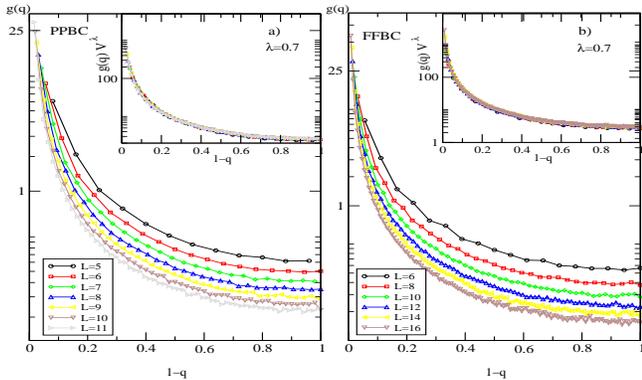}}
\caption{$g(q)$ versus $1-q$ for the PP (left panel) and FF case
(right panel) for different lattice sizes $L=5-11$ (PP) and L=6-16
(FF) from top to bottom.  In both insets we plot the scaling function
$g(q) V^{\lambda}$ vs $1-q$ with $\lambda=0.7$.
\label{FIG1}}
\end{center}
\end{figure}
\vskip -0.15in
Several numerical works have recently searched for low-lying
excitations in spin glasses using heuristic algorithms
\cite{MARTIN}. But, to our knowledge, no study has ever presented
exact results about the statistics of lowest excitations. In this
paper we have exactly computed ground states and lowest excitations in
two-dimensional Gaussian spin glasses defined by

\be
{\cal H}=-\sum_{i<j}J_{ij}\,\s_i\,\s_j
\label{eq0}
\ee

\noindent
where the $\s_i$ are the spins ($\pm 1)$ and the $J_{ij}$ are quenched
random variables extracted from a Gaussian distribution of zero mean
and unit variance.  These have been computed by using a transfer
matrix method working in the spin basis. Representing each spins state
by a weight and a graduation in the energy we can build explicitly the
ground state by keeping the largest energy and next by iteration the
first excitation and so on. The continuous values for the couplings
assures that there is no accidental degeneracy in the system (apart
from the trivial time-reversal symmetry $\sigma\to -\sigma$).
Calculations have been done in systems with free boundary conditions
in both directions (FF), periodic boundary conditions in both
directions (PP) and free boundary conditions in one direction but
periodic in the other (FP).  In all cases we find the same qualitative
and quantitative results indicating that we are seeing the correct
critical behavior.

We found ground states and lowest excitations for systems ranging from
$L=4$ up to $L=11$ for PP and up to $L=16$ for FP and FF. The number
of samples is very large, typically $10^6$ for all sizes. This
requires a big amount of computational time and calculations were done
in a PC cluster during several months. For each sample we have
evaluated the volume of the excitation $v$ (and hence the overlap
$q=1-2v/V$ between the ground state and the first excitation) and the
gap $E$. From these quantities we directly obtain $g_v$ and
$\hat{P}_v(E)$. In figure \ref{FIG1} we show
$g(q)=\frac{V}{2}g_v$ as function of $q$ for different sizes in the
PP and FF cases. We can clearly see that there are excitations of all
possible sizes but the typical ones which dominate by far are single
spin excitations. To have a rough idea of the number of rare samples
giving large scale excitations let us say that nearly half of the
total number of samples have one-spin lowest excitations, whereas less
than $10\%$ of the samples have lowest excitations with overlap $q$ in
the range $0-0.5$. This disparity increases systematically with
size. A detailed analysis of the shape of $g_v$ reveals that it has a
flat tail for large-scale excitations and a power law divergence for
finite-volume excitations.  The $g_v$ can be excellently fitted by
the following scaling form,

\be
g_v=\frac{G(q)}{V^{\lambda+1}}=\frac{1}{V^{\lambda+1}}\bigl(A+\frac{B}{(1-q)
^{\lambda+1}}\bigr)~~~.
\label{eq4}
\ee

This type of scaling, applied only to large-scale excitations, was
proposed in \cite{RS}. Although we do not have a simple explanation
for (\ref{eq4}) we emphasize how such expression interpolates
extremely well the whole spectrum of sizes matching the two volume
sectors: a small-scale sector (finite $v$) $g_v\sim 1/v^{\lambda+1}$
and a large-scale sector ($v/V$ finite) $g_v\sim
1/V^{\lambda+1}$. This matching explains why there is a unique
exponent $\lambda$ in (\ref{eq4}) which describes two completely
different volume sectors. Note that although the $g_v$ is defined for
discrete volumes, in the limit $V\gg 1$ values of $q$ are equally
spaced by $\Delta q=2/V$ and the function $g(q)=\frac{V}{2}g_v$
becomes continuous if expressed as function of $q$ instead of the
integer variable $v$,

\be
g(q)=\frac{1}{2V^{\lambda}}\bigl(A+\frac{B}{(1-q)^{\lambda+1}}\bigr)~~~.   
\label{eq5}
\ee 

Although eq.(\ref{eq4}) diverges for $q=1$ apparently leading to a
violation of the normalization condition for the $g_v$, it must be
emphasized that no excitation has $q=1$ so there is a maximum cutoff
value $q^*=1-2/V$ corresponding to one-spin excitations. The
normalization condition for the $g(q)$ using (\ref{eq5}) yields in the
large $V$ limit,

\be
\int_0^{q^*=1-2/V}g(q)dq=1\rightarrow
\frac{A-B/\lambda}{2V^{\lambda}}+\frac{B}{2^{\lambda+1}\lambda}=1
\label{eq6}
\ee

\noindent
implying $\lambda\ge 0$. The divergent term ($q\to 1$) in (\ref{eq5})
shows that one-spin excitations dominate the whole spectrum. In fact,
$g(1)\simeq O(1) \gg g(V/2)\simeq 1/V^{\lambda+1}$ so the majority of
excitations are finite-volume excitations. But the average
excitation volume $\overline v=\sum_{v=1}^V v g_v\simeq
AV^{1-\lambda}$ diverges in the $V\to \infty$ limit and differs from
the typical excitation volume $v_{\rm typ}\sim 1$. This yields an
independent measurement of the exponent $\lambda$. By fitting the
average volume to the expression $\overline v=C_1+C_2V^{1-\lambda}$ or
by measuring the ratio $g(V/2)/g(1)\sim D_1+D_2V^{-1-\lambda}$ we get
an effective exponent $\lambda_{\rm eff}=0.7\pm 0.05$ as best fitting
value. Although we find some systematic corrections in the value of
$\lambda$, our results are compatible with the general equation
(\ref{eq4}) plus some systematic subleading corrections.  We will see
later how we can obtain a better estimate of $\lambda$.
\begin{figure}[tbp]
\begin{center}
\rotatebox{270}{
\includegraphics*[width=5cm,height=8cm]{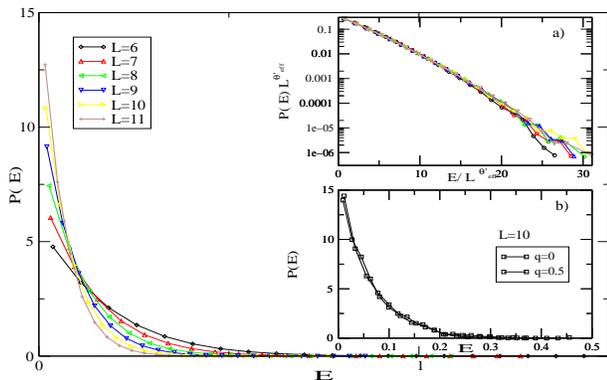}}
\caption{Gap distribution $P(E)$ vs $E$ for different
lattice sizes in the PP case. In inset a) Scaling obtained
from the ansatz (\ref{eq7}) with $\theta'_{\rm eff}=-1.6$. In inset b) we
show the $\hat{P}_v(E)$ for different
excitation sizes ($q=0.5, q=0$) for a lattice size $L=10$. Note 
that the distribution is independent of the size of the excitation.} 
\label{FIG2}
\end{center}
\end{figure}
\vskip -0.15in
After having discussed the $g_v$ we show the results for the energy
gap distribution $\hat{P}_v(E)$. In figure~\ref{FIG2} we show this
distribution for the PP case. Similar results are obtained for the FF
and FP cases. Quite remarkably, this distribution does not depend on
the size $v$ of the excitation, hence both large and small-scale
excitations are described by the same gap distribution (inset b) of
figure~\ref{FIG2}). The normalized $\hat{P}_v(E)$ has the following
scaling behavior,

\be
\hat{P}_v(E)=L^{-\theta'} {\cal P}\Bigl (\frac{E}{L^{\theta'}}\Bigr)~~~. 
\label{eq7}
\ee

Since both the exponent $\theta'$ and the scaling function ${\cal P}$
are independent of $v$ this implies that the gap probability
distribution $P(E)=\sum_{v\ge 1} g_v\hat{P}_v(E)$ satisfies the same
scaling behavior (\ref{eq7}). In the inset a) of figure~\ref{FIG2} we
also show the best data collapse for $P(E)$ obtained with an effective
exponent $\theta'_{\rm eff}\simeq -1.6$. A detailed study of the
moments of $P(E)$ computed for different values of $L$ shows that
there are also strong sub dominant corrections to the
leading scaling (\ref{eq7}) and that the value which gives the best
data collapse turns out to be $\theta'=-1.7\pm 0.1$. We will argue
below that, for Gaussian couplings, $\theta' \le -2$
and explain why finite-size effects are so big.

\begin{figure}[tbp]
\begin{center}
{
\includegraphics*[width=7.5cm,height=6cm]{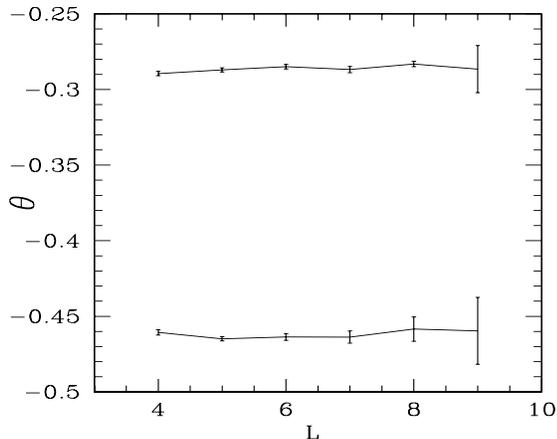}
}
\caption{Domain-wall exponent (top) and $\theta$ exponent (bottom)
estimated as explained in the text and plotted as a function of $L$.} 
\label{fig4}
\end{center}
\end{figure}

Now we want to show how the exponents $\lambda$ and $\theta'$ combine
to give the usual scaling exponent $\theta$ describing the energy cost
of {\em typical} thermal excitations in droplet theory.  One of the
most relevant results from the ansatz in (\ref{eq4}) is that both
small and large scale excitations contribute to low-temperature
properties. In general, let us consider any expression (such as (\ref{eq3}))
involving a sum over all possible volume excitations. Restricting the
sum to the large-scale regime ($v/V$ finite) the net contribution to
such sum is proportional to $Vg_V\hat{P}_{V}(0)\propto
L^{-\theta'-d\lambda}{\cal P}(0)$.  Coming back to (\ref{eq3}) we
note, using (\ref{eq4}), that both small and large-scale excitations
yield a contribution of the same order $TL^{-\theta'-d\lambda}{\cal
P}(0)$. This is a consequence of the aforementioned fact that there is
a single exponent $\lambda$ describing the large and small sectors of
the volume spectrum (\ref{eq4}). These considerations lead to the
existence of an exponent $\theta=\theta'+d\lambda$ determining the
zero-temperature critical behavior of the order parameter
$\overline{<q^2>}$.  Because of the systematic corrections in the
values of the exponents $\lambda$ and $\theta'$, in principle it is
difficult to give an accurate estimate for the exponent
$\theta$. Nevertheless, despite of these strong corrections, the
combination $\theta=\theta'+d\lambda$ seems to be very stable (see
figure~(\ref{fig4})). Thus we computed, for each size, the combination
\be
A(L)=L^2 {\overline{E}(L) \over \overline{v}(L)} \; .
\ee
Since $\overline{E}(L) \simeq L^{\theta'}$ and $\overline{v}(L)
\simeq L^{d (1-\lambda)}$, then the fit of $A(L)$ gives a direct
estimate of $\theta(L)=\theta'(L)+d\lambda(L)$. In Fig.~\ref{fig4}, we
show the effective exponent form a fit of $A(L)$ vs. $L$, with data in
the range $[L,\cdots,L_{max}=11]$ for PPBC. We also show for
comparison the effective exponent for $\theta_{DW}$ obtained also
with PPBC.  Our best values for $\theta$ is $=-0.46(1)$ for the
PPBC.  This value is very close to the finite-temperature (Monte Carlo
or transfer matrix) estimates $\theta=-0.48(1)$ \cite{JAP} but
certainly smaller than domain-wall calculations $\theta_{DW}=-0.285$
\cite{VARIOS}. Since our estimate for $\theta$ has been obtained with
a new independent method it clearly shows that $\theta\ne
\theta_{DW}$.

Now we come back to the issue why in the large volume limit $\theta'$
should converge to a value $\le -d$. The argument\cite{HUSE} goes as
follows. Consider the ground state and all possible one-spin
excitations. Because one-spin excitations are not necessarily the
lowest excitations the statistics of the lowest one-spin excitations
must yield a upper bound $\theta'_{1}$ for the value of $\theta'$,
i.e.  $\theta'\le \theta'_{1}$. The statistics of the lowest one-spin
excitations is determined by the behavior of the ground-state local
field distribution $P(h)$ in the limit $h\to 0$. If $P(h)$ is
self-averaging and $P(0)$ is finite in the large size limit then the
statistics of the lowest excitations must be governed by the exponent
$\theta'_{1}=-d$. Numerical results \cite{MF} in 2 dimensions show
that the aforementioned conditions are clearly satisfied indicating
that $\theta'\le -2$. Indeed, a determination of the finite-size
corrections to the extreme statistics of lowest one-spin excitations
shows that this asymptotic value can reached only for much larger
sizes. The disagreement between our estimated effective value
$\theta'_{\rm eff}\simeq -1.7$ in the scaling plot of figure $2$ and
the actually correct value $\theta'\le -2$ must be seen as a warning
when extracting exponents from the average behavior of extreme values
of statistical distributions for small systems.

If the power law distribution (\ref{eq4}) properly interpolates the
large and small volume sectors and then if we assume that $\theta'=-d$
holds in any finite dimension $d$, then the thermal exponent $\theta$
is given by the interesting relation $\theta=d(\lambda(d)-1)$
indicating that $\lambda(d_l)=1$ corresponds to the lower critical
dimension $d_l$. Above $d_l$ the average lowest excitation volume is
finite and $\lambda\ge 1$. Our results may open the way to properly
characterize the value of the thermal exponent $\theta$ in three
dimensions without need to use the MacMillan assumption about how to
generate low-lying typical large scale excitations. 
Preliminary results in three dimensions \cite{MF} reveal that
(\ref{eq4}) describes well the data for small sizes.
\vskip -0.15in
We have shown that a proper description of low-temperature properties
in two-dimensional Gaussian spin glasses must be done in terms of two
exponents: one ($\lambda$) describing the relative number of
finite-volume excitations compared to large scale ones, the other
($\theta'$) describing the typical energy cost of these lowest
excitations whatever their size. Assuming that $\theta'=-d$ one can
conclude that the {\em entropic} exponent $\lambda$ fully
characterizes the spin-glass phase. Although independent numerical
estimates of $\theta'$ and $\lambda$ show strong finite-volume
corrections, the thermal exponent $\theta=\theta'+d\lambda$ is quite
smooth with $L$ giving $\theta=-0.46\pm 0.01$ unambiguously showing
(given the high numerical precision for the value of $\theta_{DW}$)
that $\theta< \theta_{DW}$. To sum up, MacMillan excitations
are not the typical low-lying excitations and  our approach offers
a new and independent way to estimate the thermal exponent $\theta$
without the need to generate typical low-lying excitations.  Although
we have focused our research in the two-dimensional Gaussian spin
glass we believe that our conclusions remain valid for general
dimensions beyond $d=2$.

{\bf Acknowledgments.} 
We are grateful to D. Huse for calling us the attention on the general
validity of the result $\theta'=-d$. F.R. and M.S. are supported by
the Spanish Ministerio de Ciencia y Tecnolog\'{\i}a, project PB97-0971
and grant AP98-36523875 respectively. M. P. and F. R. acknowledge
support from the French-Spanish collaboration (Picasso program and
Acciones Integradas HF1998-0097).

\end{multicols}
\end{document}